\documentclass{aa}  

\usepackage{graphicx}
\usepackage{txfonts}
\usepackage{soul} 
\usepackage{color} 
\begin{document}

   \title{Fossil group origins: }

   \subtitle{VIII RXJ075243.6+455653 a transitionary fossil group}

   \author{J. A. L. Aguerri\inst{1,2}, A. Longobardi\inst{3,4}, S. Zarattini\inst{5,6}, A. Kundert\inst{7}, E. D'Onghia\inst{7} and L. Dom\'{\i}nguez-Palmero\inst{1,8}
        }
   \institute{Instituto de Astrof\'{\i}sica de Canarias; C/ V\'{\i}a L\'actea s/n, 38200, La Laguna, Spain
              \email{jalfonso@iac.es}
              \and 
              Departamento de Astrof\'{\i}sica, Universidad de La Laguna, E-38206 La Laguna, Spain
              \and
              Kavli Institute for Astronomy and Astrophysics, Peking University, 5 Yiheyuan Road, Haidian District, Beijing, 100871, PR China
              \and
              Max-Planck-Institut f\"ur Extraterrestrische Physik, Giessenbachstrasse, D-85741 Garching, Germany
              \and
              Dipartimento di Fisica, Universit\`a degli Studi di Trieste, via Tiepolo 11, I-34143 Trieste, Italy
              \and 
               INAF-Osservatorio Astronomico di Trieste, via Tiepolo 11, I-34143 Trieste, Italy
               \and
               Department of Astronomy, University of Wisconsin-Madison, 475 N. Charter St., Madison, WI 53706, USA
               \and
               Isaac Newton Group of Telescopes, Apartado 321, E-38700 Santa Cruz de La Palma, Canary Islands, Spain
             }


 
  \abstract
   {It is thought that fossil systems are relics of  structure formation in the primitive Universe. They are galaxy aggregations that have assembled their mass at high redshift with few or no subsequent accretion. Observationally these systems are selected by large magnitude gaps between their 1st and 2nd ranked galaxies ($\Delta m_{12}$). Nevertheless, there is still debate over whether or not this observational criterium selects dynamically evolved ancient systems.}
   {We have studied the properties of the nearby fossil group RXJ075243.6+455653 in order to understand the mass assembly of this system.}
   {Deep spectroscopic observations allow us to construct the galaxy luminosity function (LF) of RXJ075243.6+455653 down to $M_{r}^{*}+ 6$. The analysis of the faint-end of the LF in groups and clusters provides valuable information about the mass assembly of the system. In addition, we have analyzed the nearby large-scale structure around this group.}
   {We identified 26 group members within $r_{200} \sim 0.96$ Mpc. These galaxies are located at $V_{c} = 15551 \pm 65$ km s$ ^{-1}$ and have a velocity dispersion of $\sigma_{c} = 333 \pm 46$ km s$^{-1}$. The X-ray luminosity of the group is $L_{X} = 2.2 \times 10^{43}\ h_{70}^{-2}$ erg s$^{-1}$, resulting in a mass of M$ = 4.2 \times 10 ^{13}\ h_{70}^{-1}$ M$_{\odot}$ within 0.5$r_{200}$. The group has $\Delta m_{12} = 2.1$ within 0.5$r_{200}$, confirming the fossil nature of this system. RXJ075243.6+455653 has a central brightest group galaxy (BGG) with $M_{r} = -22.67$, one of the faintest BGGs observed in fossil systems. The LF  of the group shows a flat faint-end slope ($\alpha = -1.08 \pm 0.33$). This low density of dwarf galaxies is confirmed by the low value  of the dwarf-to-giant ratio ($DGR = 0.99\pm0.49$) for this system.  Both the lack of dwarf galaxies and the low luminosity of the BGG suggests that RXJ075243.6+455653 still has to accrete mass from its nearby environment. This mass accretion will be achieved because it is the dominant structure of a rich environment formed by several groups of galaxies (15) within $\sim 7$ Mpc from the group center and with $\pm 1000$ km s$^{-1}$.}
   {RXJ075243.6+455653 is a group of galaxies that has not yet completed the process of its mass assembly. This new mass accretion will change the fossil state of the group. This group is an example of a galaxy aggregation selected by a large magnitude gap but still in the process of the accretion of its mass. }

   
   \authorrunning{J. A. L. Aguerri et al.}
   \titlerunning{RXJ075243.6+455653 a transitionary fossil group}

   \maketitle
%

\section{Introduction}

The structure formation in the Universe is driven by mergers and accretion of small substructures into larger ones \citep[][]{white1978}, and these processes are still active at present. Nevertheless, a fraction of the structures present today could be relics of the primitive Universe. These systems are called fossil groups (FG) or, more generically, fossil systems (FS). These ancient galaxy aggregations were formed at high redshift ($z>1$) with few or no subsequent accretion \citep[][]{donghia2005}.  But, do these objects really exist today?; are there systems in the Universe for which the hierarchical formation is frozen since several Gyr ago? If this kind of system exists, they are important because they contain valuable information about the assembly of massive halos in the primitive Universe. We can therefore understand the formation processes that occurred several Gyr ago by analyzing the properties of nearby FSs.

\cite{ponman1994} established that the group RXJ1340.6+4018 was the remnant of an ancient system of galaxies.  This system shows extended X-ray emission, indicating that the group is embedded in a massive dark matter halo. In addition, its light is dominated by a central early-type galaxy. The 2nd ranked luminous galaxy belonging to the group is more than 2 magnitudes fainter than the brightest one. The lack of $L^{*}$ galaxies in the system was interpreted as the result of the central galaxy cannibalizing the surrounding luminous objects over time. These characteristics made  this system a candidate for an early-formed and relaxed structure, and it was classified as a fossil relic.

After the discovery of RXJ1340.6+4018, other systems were observed with  large magnitude gaps between their 1st and 2nd ranked galaxies ($\Delta m_{12}$). The presence of a large $\Delta m_{12}$ in those systems was interpreted as indication of fossilness. \cite{jones2003} gave the first observational definition of FSs as those galaxy aggregations with $\Delta m_{12} > 2.0$ mag in the $r$-band for the galaxies of the systems within half its virial radius. In addition, the central galaxy should be surrounded by an extended diffuse X-ray halo with $L_{X} > 10^{42} h_{50}^{-2}$ erg s$^{-1}$. This definition was slightly modified by \cite{dariush2010} according to the analysis of cosmological simulations. They claim that the magnitude gap between the 1st and 4th ranked galaxies of the system ($\Delta m_{1,4}$) was a better quantity to select old and dynamically relaxed galaxy aggregations. In particular, they proposed that fossil systems are those with $\Delta m_{1,4} > 2.5$ for galaxies within half of their virial radius.

Several samples containing fossil candidates have been selected during the last decade in different galaxy surveys based on large values of their $\Delta m_{12}$ \citep[e.g.,][]{khosroshahi2007, santos2007, voevodkin2010, proctor2011, harrison2012}. Many works have obtained the observational properties of these systems. These properties can be grouped in (i) properties of the hot intracluster component; (ii) properties of the central dominant galaxy or the brightest group galaxy (hereafter BGG); and (iii) properties of the galaxy satellite population.

The global X-ray scaling relations of dark matter halos are related to their formation. Halos formed early should have some imprints in their X-ray properties. In particular, more concentrated dark matter halos collapse first in the hierarchical scenario, which can produce a greater compression of the gas in the central regions of the halo and enhance the X-ray temperature ($T_{X}$) and luminosity ($L_{X}$). Some works suggest that X-ray global scaling relations in FS and non-FS are different. In particular, FSs were found to have larger $T_{X}$ than non-fossil ones \citep[e.g.,][]{khosroshahi2007}. These differences between FS and non-FS were also reported in scaling relations combining optical and X-ray quantities, such as the $L_{X} - L_{opt}$ or $L_{X} - \sigma$ relations \citep[e.g.,][]{proctor2011, khosroshahi2014};  they  were interpreted as an indication that the halos of FSs formed early in the high redshift Universe. Nevertheless, larger samples of FSs and non-FSs have obtained that those differences do not exist, indicating that according to the global scaling relations, fossil and non-fossil halos are similar \citep[e.g.,][]{harrison2012, girardi2014, kundert2015}.

The expected early formation in FSs, with no later accretion, should also produce some peculiarities in the observational properties of their BGGs. In particular, their formation at high redshift could be driven by mergers with large amounts of gas (wet mergers). \cite{khosroshahi2006} reported differences in the isophotes shape between the BGGs in FS and non-FS pointing towards a formation via wet mergers. They obtained that BGGs of FSs have more disky isophotes than those in non-FSs. Nevertheless, these differences in the isophotal shape were not confirmed by other samples of galaxies \citep[][]{mendezabreu2012}. Indeed, the scaling relations of BGGs in FSs show that they assembled a fraction of their mass by wet mergers probably at high redshift. However, a significant fraction of their mass was accreted via later dry mergers \citep[][]{mendezabreu2012}. The similar properties of the stellar populations of BGGs in FSs and non-FSs also point toward a similar formation \citep[][]{labarbera2009, eigenthaler2013, trevisan2017}. Nevertheless, the merging process in the BGGs of FSs has been especially effective because these galaxies  are among the brightest galaxies observed in the Universe \citep[][]{aguerri2011, mendezabreu2012, zarattini2014}.  

The galaxy population of FSs and non-FSs are clearly different in the bright-end of the galaxy luminosity function (LF). In particular, FSs show a lack of $L^{*}$ galaxies \citep[e.g.,][]{khosroshahi2006, mendesdeoliveira2006, aguerri2011, adami2012, lieder2013, khosroshahi2014, zarattini2014}. Additionally, there is also evidence for differences in the number of low-mass galaxies - FSs show LFs with shallower faint-end than non-FSs \citep[][]{zarattini2015}. This lack of substructure in fossil halos has been interpreted as a challenge for the hierarchical formation \citep[][]{donghia2004}; nevertheless, this is still a puzzle and a matter of debate \citep[][]{zibetti2009}. The galaxy population in FSs also shows evidence of substructure similar to non-FSs, arguing against the relaxed and old dynamical age of FSs \citep[][]{zarattini2016}.

The previously reported properties indicate that selection of systems by only a large magnitude gap does not guarantee the selection of galaxy aggregations that are dynamically old, early forming, and with few subsequent accretions. This has also been pointed out by some numerical cosmological simulations that analyzed the properties of systems with large magnitude gaps. \cite{vonbendabeckman2008} showed that the fossilness of a system is a transition phase, whereby galaxy groups and clusters could pass several times through a fossil phase along their evolution. Recently, \cite{kundert2017} found that, on average, the fossilness of a system changes over 2 - 3 Gyr. Some simulations also show that  the last major merger in BGGs in FSs has occurred more recently than in similar galaxies in non-FSs \citep[][]{diazgimenez2008, kundert2017}. The orbital structure of the galaxies in FSs increases the efficiency of mergers into the BGGs in FSs more than in non-FSs \citep[][]{sommerlarsen2006}. This produces FSs BGGs that are more massive on average than those in non-FSs \citep[][]{kundert2017}. Simulations also show that there are differences at the faint-end of the LF between FSs and non-FSs. Thus, \cite{gozaliasl2014} obtained that there was almost no evolution  in the faint-end LF in FSs since $z \sim 1$. In contrast, non-FSs show a strong evolution of the faint-end. \cite{kundert2017} demonstrated that the most important difference between FSs and non-FSs is related with the halo mass accretion history over the past few Gyr. This difference in the last fraction of the accretion of the mass could be related with the surrounding large-scale environment \citep[][]{diazgimenez2011}. It is likely that a combination of the magnitude gap with other parameters provides a better selection of dynamically old systems. According to \cite{raouf2014} the combination of the magnitude gap and the luminosity of the central BGG is one of the solutions to obtain dynamically old structures. 


The present paper is part of the Fossil Group Origins (FOGO) project. This project aims at a multiwavelength characterization of a large sample of FSs \citep[see][]{aguerri2011}. We have analyzed so far the properties of their (i) dark matter halos \citep[][]{girardi2014, kundert2015}, (ii) the central BGGs \citep[][]{mendezabreu2012, zarattini2014}, and (iii) the galaxy populations \citep[][]{zarattini2015, zarattini2016}. These observational properties have been compared with state-of-the-art cosmological simulations \citep{kundert2017}. In the present work we have obtained deep spectroscopic data of the fossil group RXJ075243.6+455653 down to $\sim M_{r}^{*} +6$. We have analyzed the properties of this group focusing on the faint-end of its galaxy LF and its large-scale nearby structure.   As reported by \cite{raouf2014}, this group has the properties to be an old and evolved galaxy aggregation. In particular, it shows $\Delta m_{12} > 2.0$ and one of the faintest BGGs ($M_{r} = -22.67$) observed in FSs \citep[][]{zarattini2014}. 

The paper is organized as follows. Section 2 shows the data used. The results are presented in Sect. 3. The discussion and conclusions are given in Sects. 4 and 5, respectively. Through this paper we have used the cosmology given by $H_{0} = 70$ km s$^{-1}$ Mpc$^{-1}$, $\Omega_{\Lambda} = 0.7$, and $\Omega_{m} = 0.3$. Using this cosmology, RXJ075243.6+455653 is located at a distance of 230.6 Mpc, resulting in a scale of 1.01 kpc arcsec$^{-1}$.

\section{The data for RXJ075243.6+455653}

We downloaded the photometric and velocity information for galaxy selected objects from Sloan Digital Sky Survey Data Release 6 \citep[SDSS-DR6;][]{adelman2008} within a radius of 4 Mpc around the coordinates $\alpha (J2000)= 07^{h}52^{m}44.2^{s}$ and $\delta (J2000) = +45^{o}56^{'}57.4^{''}$. These central coordinates correspond to the peak of X-ray emission detected from the ROSAT satellite and are associated with the group RXJ075243.6+455653. This central X-ray peak has a luminosity of $L_{X} = 2.20 \times 10^{43}$ erg s$^{-1}$ \cite[][]{zarattini2014}.

The catalog turned out to have  89 galaxies brighter than m$_{r} \sim 18.0$ with measured radial velocities. Using this data-set from SDSS-DR6, \cite{zarattini2016} reported that the mean velocity of group ($V_{c}$) was 15498 km s$^{-1}$, and the velocity dispersion $\sigma_{c} = 259$ km s$^{-1}$. The estimate of $L_{X}$ allowed us to calculate $r_{200} = 0.96$ Mpc \citep{zarattini2014} using the relations from \cite{bohringer2007} and \cite{arnaud2005}. The mass of the system was obtained following \cite{girardi1998} and \cite{girardi2001} resulting in $M = 4.2 \times 10^{13} M_{\odot}$ \citep[][]{zarattini2014}.

The photometric catalog acquired from SDSS-DR6 was used  to select targets for additional multi-object spectroscopy with different instruments. In particular, we selected those galaxies with $g - r < 1.1$, $m_{r} < 19.5$ and within 0.5$^{o}$ ($\sim$ 1.8 Mpc radius at the distance of the group) from the group center to be observed with the AF2@WHT instrument. These objects  will be referred to as spectroscopic targets. A total of 129 spectroscopic targets were selected and located in two AF2 fiber configurations. We obtained low-resolution spectra ($R = 280$ and grism R158B) in three exposures of 1800s per pointing, with the spectra reaching $S/N > 5$ for a secure redshift determination \citep[see e.g.,][]{agulli2016}. We also selected targets with $g - r <1.1$ and $19.5 < m_{r} < 21.0$  to be observed with the OSIRIS@GTC instrument. A total of 121 spectroscopic targets were located in nine  different OSIRIS pointings within $r_{200}$. In this case the low-resolution  spectra (R=360) of these objects were obtained using the grism R300B in three exposures of 1200s for each pointing. These new observations consisted of a total of 250 new spectra.

The new spectroscopic data was reduced using different pipelines. The reduction of the data from AF2@WHT was achieved with the instrument pipeline, version 3.0 \citep[see][]{dominguezpalmero2014}. In contrast, the data from OSIRIS@GTC was reduced by using the GTCMOS pipeline\footnote{http://www.inaoep.mx/$\sim$ydm/gtcmos/mos\_reduction.html}.

\subsection{Velocity determination and spectroscopic completeness}

The recessional velocities of the new spectra were obtained using rvsao.xcsao IRAF task \citep[see][]{kurtz1992}. This task cross-correlates a template spectrum library with the observed galaxy spectrum. In this case we have used as template library the one given by \cite{kennicutt1992}.  A total of 191 new velocities were determined by this technique. This makes a total of 280 galaxies with measured radial velocities within 4 Mpc from the group center.

\begin{figure}
        \centering
        \includegraphics[width=\hsize]{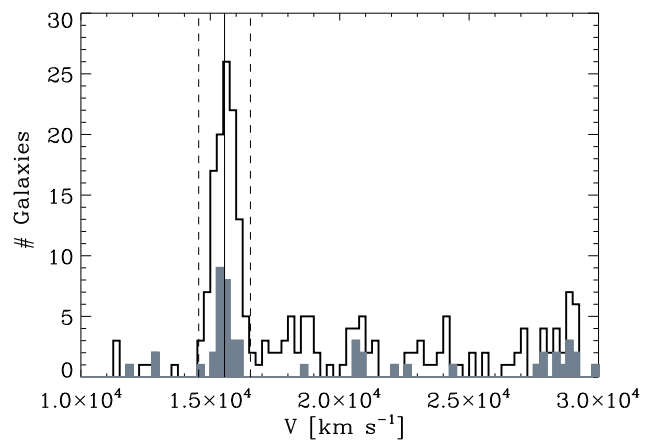}
        \caption{Velocity histogram of the galaxies in the direction of RXJ075243.6+455653. The black histogram shows the radial velocities of the galaxies within 4 Mpc from the cluster center. The gray filled histogram corresponds to galaxies within $r_{200}$. The vertical line shows the mean velocity of the group. The dashed vertical lines are located at $\pm 3 \sigma_{c}$ from $V_{c}$.}
         \label{vhisto}%
         \end{figure}

Figure \ref{vhisto} shows the velocity histogram of the galaxies in the direction of RXJ075243.6+455653 and within an aperture of 4 Mpc. A prominent peak can be observed at $V_{c} \sim 15550$ km s$^{-1}$ which corresponds to the mean velocity of the group \citep[see][]{zarattini2014}. The velocity histogram also shows that the peak in velocity corresponding to the group is not isolated. In contrast, there is a continuum of background galaxies at larger velocities. The lack of isolation of this group can also be seen by the large number of galaxies with velocities $V_{c} \pm 3\sigma_{c}$ out from $r_{200}$ (see Fig. \ref{vhisto}). This suggest that the group is surrounded by other galaxies or groups.

The group members were determined by using a simple $\sigma-$clipping algorithm. Thus, we computed several times the mean velocity ($V_{c}$) and velocity dispersion $\sigma_{c}$ by using the galaxies with recessional velocities within $V_{c} \pm 3 \times \sigma_{c}$ and inside $r_{200}$. This iterative process stopped when no changes in $V_{c}$ and $\sigma_{c}$ from one step to the other was obtained. This process returns a total of 26 groups members with $m_{r} < 21.0$ and within $r_{200}$. Using these group members we obtain: $V_{c} = 15551 \pm 65$ km s$^{-1}$ and $\sigma_{c} = 333 \pm 46$ km s$^{-1}$. The remaining 254 objects with radial velocities  out from the $V_{c} \pm 3 \times \sigma_{c}$ range and within 4 Mpc from the group center will be called background galaxies.

\begin{figure}
        \centering
        \includegraphics[width=\hsize]{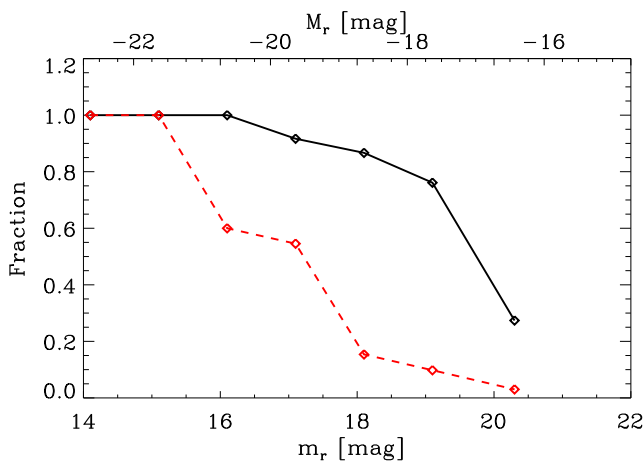}
        \caption{Spectroscopic completeness (black) and fraction of group members (red dashed) as a function of the r-band magnitude.}
         \label{completitud}%
         \end{figure}

Figure \ref{completitud} shows the spectroscopic completeness ($C$) as a function of the $r$-band apparent magnitude ($m_{r}$). This completeness was obtained by $C= N_{vel} / N_{phot}$, where $N_{vel}$ and $N_{phot}$ represents the number of galaxies with measured velocities and the number of spectroscopic targets per magnitude bin, respectively \citep[see][]{agulli2014}. In addition, Fig. \ref{completitud} also shows the fraction of cluster members ($f_{mem}$) as a function of $m_{r}$. This fraction is given by: $f_{mem} = N_{mem} / N_{vel}$, where $N_{mem}$ shows the number of galaxy group members per magnitude bin \citep[][]{agulli2014}. The spectroscopic completeness is 100$\%$ for galaxies brighter than $m_{r} =16.6$. This completeness slowly decreases to $\sim 80\%$ at $m_{r} \sim 19$, and $\sim 25\%$ at $m_{r} \sim 20.5$. The fraction of members decreases steeply with the magnitude. At $m_{r}$ fainter than 18.0, less than 15$\%$ of the observed galaxies turned out to be group members. This indicates the low efficiency at faint magnitude for the selection of group members by a simple color cut.

\begin{figure}
        \centering
        \includegraphics[width=\hsize]{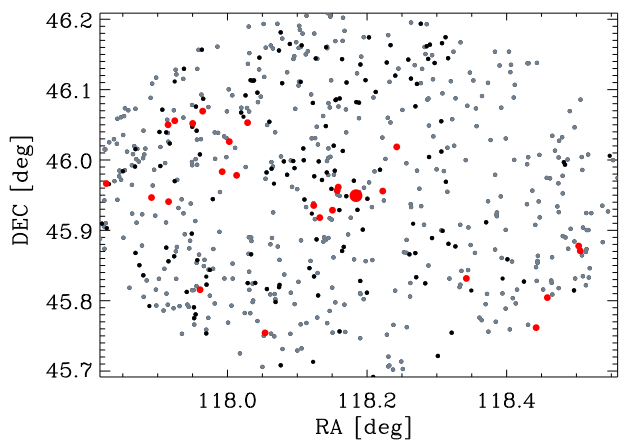}
        \caption{Sky position of the spectroscopic targets (gray), background objects (black), and galaxy group members (red) within $r_{200}$. The largest red point represents the position of the brightest galaxy in the group. }
         \label{skypos}%
         \end{figure}

Figure \ref{skypos} shows the sky distribution of the spectroscopic targets, background galaxies, and member galaxies within $r_{200}$ around the center of RXJ075243.6+455653.

\section{Results}

\subsection{The fossilness of RXJ075243.6+455653}

\begin{figure}
        \centering
        \includegraphics[width=\hsize]{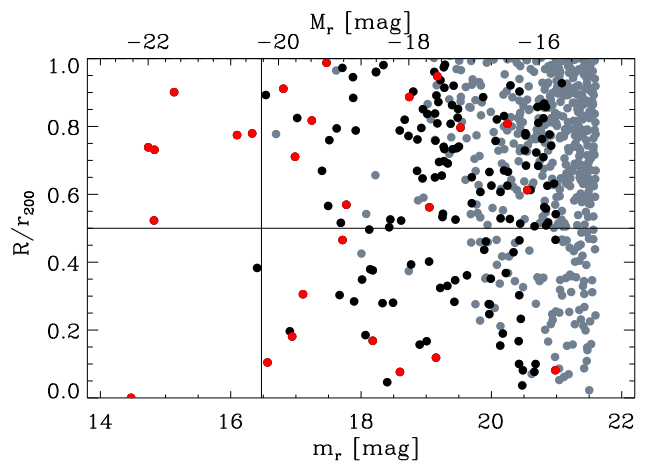}
        \caption{Magnitude-distance diagram for the group RXJ075243.6+455653. The distance to the BGG is shown as a function of the $r$-band magnitude for each galaxy within $r_{200}$. The gray, black, and red points represent the spectroscopic targets, background galaxies, and group members, respectively. The horizontal and vertical full lines represent $R/r_{200} = 0.5$ and $\Delta m_{12} =2.0$, respectively }
         \label{magdis}%
         \end{figure}

Figure \ref{magdis} shows the distance of the galaxies from the BGG as a function of their r-band apparent magnitude. According to the definition of fossil groups given by \cite{jones2003}, RXJ075243.6+455653 is a fossil system. It shows a X-ray luminosity larger than $10^{42}$ erg s$^{-1}$, and the magnitude gap between the 1st and 2nd ranked galaxies within 0.5$R_{200}$ is $\Delta m_{12} = 2.1$. Furthermore, the magnitude gap between the 1st and 4th ranked galaxies of the system within 0.5$R_{200}$ is $\Delta m_{14}=2.55$ \citep[][]{zarattini2014}, meeting the fossil criteria of \cite{dariush2010}.

There are several group members within $0.5 < R/r_{200}< 1.0$ with magnitudes similar to the BGG. Thus, these galaxies result in $\Delta m_{12} < 2$ within $r_{200}$. We can speculate that probably these galaxies are in the process of falling into the group, and will pass closer to the group center and merge with the BGG in the next few Gyr. This  would imply that the fossilness of RXJ075243.6+455653 is a transition state as reported by some cosmological simulations \citep[][]{vonbendabeckman2008, kundert2017}. The transitory nature of the fossil state has also been reported in some other observed systems in the literature \citep[e.g.,][]{zarattini2014}
         
According to the definition of fossil systems, the fossilness strongly depends on the adopted value of $r_{200}$ of the structure. Figure \ref{magdis} shows that changes in the value of $r_{200}$ could change the fossillness of RXJ075243.6+455653. The value of $r_{200}$ for this group was determined from the X-ray luminosity obtained from RASS data. We have also computed $r_{200}$ by using  the velocity dispersion of the system and  the relation: $r_{200} =  \frac{\sqrt{3} \sigma_{c}}{10 H(z_{c})}$, where $H(z_{c})$ is the value of the Hubble constant at the redshift of the group ($z_{c}$) \citep[e.g.,][]{carlberg1997, aguerri2007}. In this case we have calculated $r_{200} = 0.80$ Mpc, which is $\sim 17 \%$ smaller than the obtained from X-ray data. This difference could be related to the luminous AGN that the central galaxy of RXJ075243.6+455653 hosts \citep[see][]{kundert2015}. Nevertheless, the fossilness of the system does not change with the new $r_{200}$ computed (see Fig. \ref{magdis}).

We can also investigate whether or not the X-ray luminosity of the AGN would affect  the fossilness definition of RXJ075243.6+455653. \cite{girardi2014} obtained a relation between the X-ray luminosity and the optical luminosity ($L_{opt}$) for a sample of clusters and fossil systems. According to this work, RXJ075243.6+455653 shows an optical luminosity of $L_{opt} = 3.05 \times 10^{11}\ h_{70}^{-2}$ L$_{\odot}$. Taking into account the Girardi's relation, RXJ075243.6+455653 would have a X-ray luminosity of $L_{X}=5.78 \times 10^{42}\ h_{70}^{-2}$ erg s$^{-1}$. This value is larger than 10$^{42}\ h_{50}^{-2}$ erg s$^{-1}$ chosen as  limit to define fossil systems. Therefore, we can say that the fossillness of RXJ075243.6+455653 is not compromised by the X-ray luminosity of the central AGN. We also note that the limit in the X-ray luminosity used to define fossil systems is taken in order to avoid isolated galaxies. In this case, the velocity dispersion of the system shows us that RXJ075243.6+455653 is a group-like halo.

\subsection{The color-magnitude diagram}

\begin{figure}
        \centering
        \includegraphics[width=\hsize]{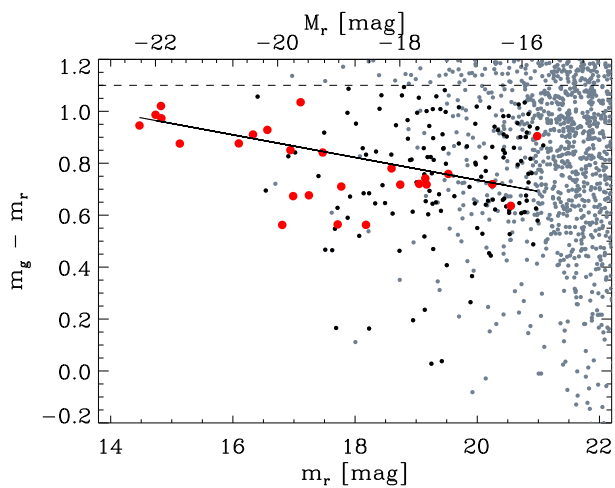}
        \caption{Color-magnitude diagram of the galaxies within $r_{200}$ from the group center. The gray points represent all the galaxies in the direction of the group. Red and black points show the group members and the background galaxies, respectively. The horizontal dashed line represents the color limit chosen for the selection of the spectroscopic targets. The full line shows the fitted red sequence.}
         \label{colormag}%
         \end{figure}

Figure \ref{colormag} shows the color-magnitude diagram of the galaxies located within $r_{200}$ around the center of RXJ075243.6+455653. We have fit the red sequence (RS) for those group members with $m_{r} < 18.0$ and with $g - r > 0.8$. The best fit is $(g - r)_{RS} = -0.043 \pm 0.01 m_{r} +1.606 \pm 0.20$ and $\sigma_{RS}=0.036$. Taking into account the errors of the fit of the red sequence, the intrinsic scatter is $\sigma_{RS,i} = 0.16$. The red sequence of the group is well defined down to the dwarf regime ($M_{r} \sim -16.0$). 

We have divided the group members into red and blue galaxies. We have considered blue galaxies as those group members with $g - r < (g - r)_{RS} - 3\sigma_{RS}$, where $(g - r)_{RS}$ and $\sigma_{RS}$ are the color and the dispersion of the red sequence \citep[see e.g.,][]{agulli2014}. We have obtained six groups members classified as blue galaxies, implying that the blue fraction of this group is 0.23. These blue galaxies are located in the external regions of the group, at a mean distance of 0.7 $r_{200}$. This number of blue galaxies can be considered as an upper limit. No blue galaxies are obtained if we consider blue galaxias as those group members with $g - r < (g - r)_{RS} - 3\sigma_{RS,i}$.

We also note the absence of blue galaxies fainter than $\sim M_{r}=-18.0$. Although some blue galaxies have been spectroscopically observed in this magnitude range, they turned out be background objects. Due to the small number of observed galaxies, it is difficult to be certain of a lack of blue dwarf galaxies in this group.

\subsection{The spectroscopic galaxy luminosity function}

\begin{figure}
        \centering
        \includegraphics[width=\hsize]{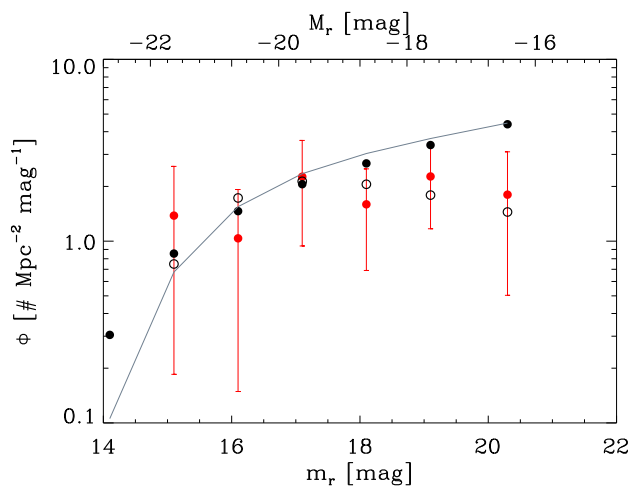}
        \caption{Luminosity function of the galaxies within $r_{200}$ from the group center (red points). The Schechter fits of the composite luminosity functions for systems with $\Delta m_{12} < 0.5$ and $\Delta m_{12} > 1.5$ from \cite{zarattini2015} are also overplotted with filled and empty black circles, respectively.  The gray full line represents the Schechter fit of the LF for groups with $log(M) = 13.56$ from \cite{zandivarez2011}. All LFs have been normalized in the magnitude interval 14 $< m_{r} < 18$.}
         \label{lumfun}%
         \end{figure}

Figure \ref{lumfun} shows the spectroscopic galaxy LF  of the group within $r_{200}$. This LF was computed as $\Phi(M_{r}) = N_{phot}(M_{r}) \times f_{mem}(M_{r}) / A$, where A is the area of the observations, $N_{phot}$ is the number of photometric targets per magnitude bin, and $f_{mem}$ is the fraction of group members per magnitude bin \citep[see e.g.,][]{agulli2014}. The uncertainties were obtained by  Poisson statistics.

Although the uncertainties of the LF are large, we can see that the faint-end is flat.  We have fit the observed spectroscopic LF by a Schechter function. The reported faint-end slope was $\alpha = -1.08 \pm 0.33$. The large uncertainty in $\alpha$ results from the large observational errors in the LF. This slope is less steep than others from massive clusters \citep[$\alpha \sim -1.5;$][]{agulli2014, agulli2016} or the field \citep[][]{blanton2005}. Nevertheless, the slope is similar within the uncertainties with the  LF in groups of galaxies \citep[$\alpha \sim -1.16$; ][]{zandivarez2011} or some dynamically young massive clusters \citep[$\alpha \sim -1.13$;][]{agulli2017}. 

The large uncertainty in the $\alpha$ parameter of the fitted Schechter function makes comparison with other samples difficult. In particular, we aim to compare the LF of the group with the composite LFs of \cite{zarattini2015} for systems with small ($\Delta m_{12} < 0.5$) and large ($\Delta m_{12} > 1.5$) magnitude gaps. To avoid this problem, we have used the Pearson test. This statistical test reports that the LF of RXJ075243.6+455653 can be modeled by the Schechter function fitted for systems with large magnitude gap ($\Delta m_{12} > 1.5 ; \chi^{2} = 0.16$). However, the LF of systems with small $\Delta m_{12}$ is not a good representation of the LF of RXJ075243.6+455653 $(\chi^{2} =23.03$). In addition, the Pearson test also suggests that the \cite{zandivarez2011} LF for systems with similar mass as RXJ075243.6+455653 is not a good representation of our data ($\chi^{2} = 10.05$). This analysis indicates that the LF of this group is similar to the LF of systems showing flat faint-end slopes, such as the systems with large magnitude gaps reported by \cite{zarattini2015}.

The flatness of the LF can also be reflected in the dwarf-to-giant ratio (DGR). This quantity measures the ratio between the number of dwarf and bright galaxies in clusters and groups. Following the definition from \cite{popesso2006}, we have considered dwarf galaxies as those members with $-18.0 < M_{r} < -16.5$, and  bright group members as those with $M_{r} < -20.0$. The obtained DGR for the members within $r_{200}$ for RXJ075243.6+455653 was 0.62, indicating that this group contains less than one dwarf galaxy per giant. 

\subsection{The large-scale environment}

We have analyzed the degree of isolation of RXJ075243.6+455653 by searching in the NASA Extragalactic Database (NED) for galaxy systems within 2 degrees of the group center; corresponding to a search radius of $\sim 7.2$ Mpc at the distance of the group.  Sixteen galaxy aggregations with radial velocity in the range $12000 < V < 18000$ km s$^{-1}$ were found in this search, with 15 of them  located within $\pm 3 \sigma_{c}$  from the central radial velocity of RXJ075243.6+455653. In addition, two of these groups are located at a projected distance from the group center smaller than $r_{200}$. The data obtained from NED are heterogeneous. Nevertheless, NED has one of the largest datasets on clusters and groups of galaxies. The 16 galaxy aggregations retrieved from the NED database come from six sources in the literature. They were obtained from data of large galaxy surveys as SDSS \citep[][]{smith2012,berlind2006,miller2005,mcconnachie2009} and the Two Micron All-Sky Survey  \citep[2MASS; ][]{diazgimenez2015,crook2007}. We have not identified groups or clusters within 2 degrees of the group center and with $V < 20000$ km s$^{-1}$ in other large cluster surveys as maxBCG \citep[][]{koester2007}, WHL \citep[][]{wen2009}, or RedMaPPer \citep[][]{rykoff2016}.

\begin{figure*}
        \centering
        \includegraphics[width=\hsize]{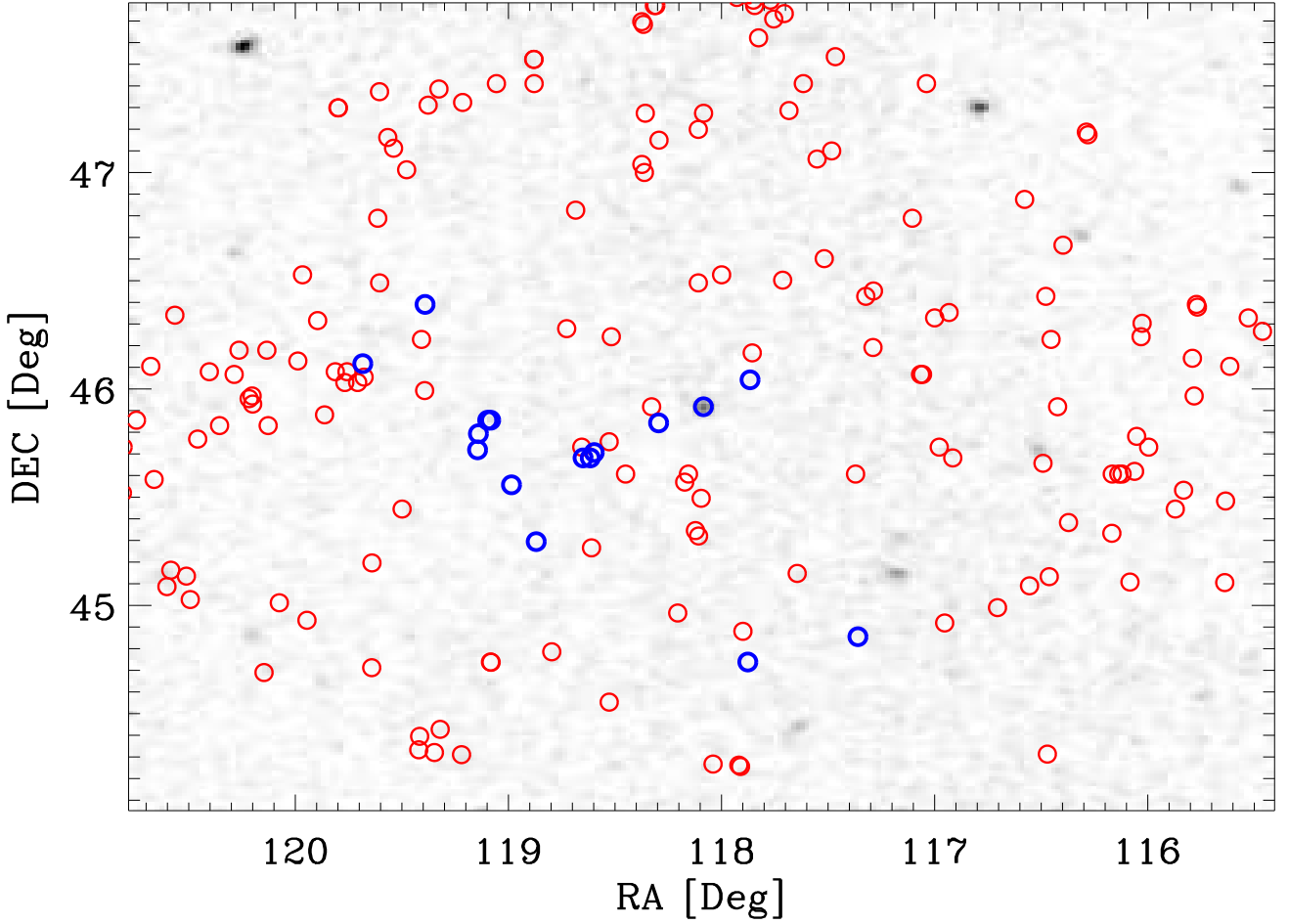}
        \caption{ROSAT image centered at the position of RXJ075243.6+455653. The open circles show the position of other groups and galaxy clusters reported by NED within 2 degrees from the group center. The red symbols represent objects with radial velocities larger than 18000 km s$^{-1}$ or smaller than 12000 km s$^{-1}$. Blue circles show groups with radial velocities in the range $12000 < V < 18000$ km s$ ^{-1}$.}
         \label{rosat}%
         \end{figure*}

\begin{figure}
        \centering
        \includegraphics[width=9cm]{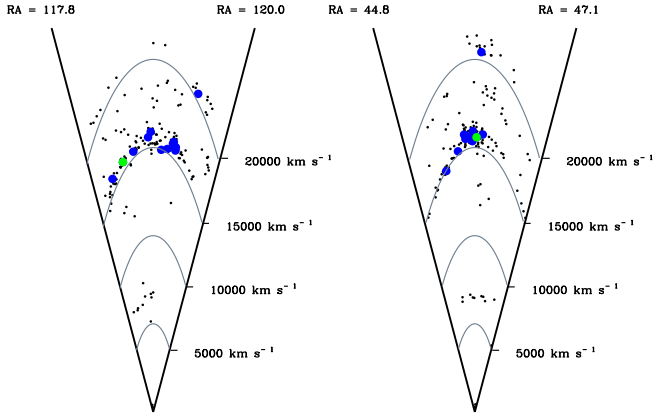}
        \caption{Conical diagrams for the RXJ075243.6+455653 group (green point) and other galaxy groups (blue points). The black points represent galaxies with $V < 22000$ km s$^{-1}$. The gray lines shows the iso-velocity curves corresponding to 5000, 10000, 15000, and 20000 km s$^{-1}$.}
         \label{conidia}%
         \end{figure}

Figure \ref{rosat} shows the position on the sky of the galaxy groups and clusters obtained from NED in a search radius of 2 degrees around the group center. The objects with radial velocities in the range $12000 < V < 18000$ km s$^{-1}$ are located in a filamentary structure which extends for several Mpc. In contrast, the background groups are more randomly distributed. Figure \ref{conidia} represents the conical diagrams with the position of RXJ075243.6+455653 and the other groups. This plot shows clearly that most of the groups have a similar radial velocity as RXJ075243.6+455653. We can conclude that RXJ075243.6+455653 is not an isolated structure. Indeed, this group forms part of a large and rich structure of groups of galaxies located within a radius of $\sim 7.2$ Mpc from the group center. The filamentary distribution of the nearby groups could indicate that they are falling into the potential well of RXJ075243.6+455653.

\begin{figure}
        \centering
        \includegraphics[width=\hsize]{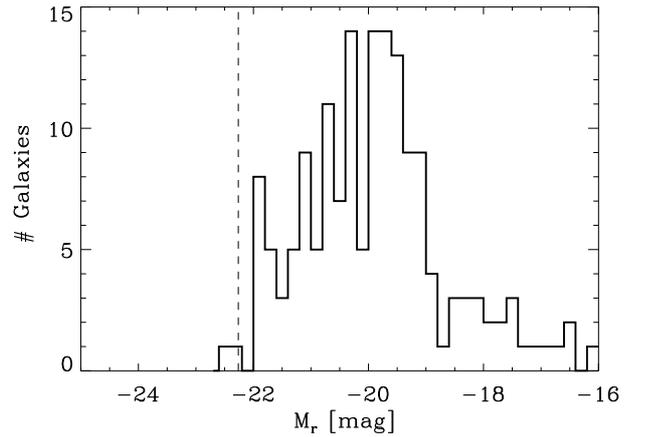}
        \caption{Absolute r-band magnitude  of the galaxies within 2 degrees from the center of RXJ075243.6+455653 and with $|V - V_{c}| <   3\sigma_{c}$. The vertical dashed line represent the absolute magnitude of the BGG of RXJ075243.6+455653.}
         \label{maghisto}%
         \end{figure}

Figure \ref{maghisto} shows the absolute magnitude histogram of the galaxies located within 2 degrees from the center of RXJ075243.6+455653 and with $|V - V_{c}| <  3\sigma_{c}$. The central galaxy of RXJ075243.6+455653 is the second brightest object in this region of the sky. There is another galaxy 0.14 mag brighter and located at 2.5 Mpc from the center of RXJ075243.6+455653. We can consider that these two galaxies are the dominant ones in the field. Figure \ref{rosat} shows the X-ray emission image from the ROSAT satellite in which it can be seen that the group RXJ075243.6+455653 is located in the X-ray peak in the center of the image. No other galaxy group from NED with $12000 < V < 18000$ km s$^{-1}$ is associated with an X-ray peak that can be observed in the ROSAT image.
 
We can speculate that due to the bright luminosity of the BGG and its position at the peak of the X-ray, the RXJ075243.6+455653 group is the dominant structure in this region of the sky. Then, the small groups of galaxies located around RXJ075243.6+455653 would contribute to the hierarchical formation of this system. In addition, the bright galaxies could merge with the central galaxy of RXJ075243.6+455653 increasing its luminosity, hence reinforcing its fossilness nature. The accretion of the different galaxy groups could also bring dwarf galaxies into the group and will increase the  faint-end slope of the LF of RXJ075243.6+455653.


\section{Discussion}

The results obtained in Sect. 3 indicate that RXJ075243.6+455653 is not the remnant of an evolved and ancient group. In contrast, this galaxy aggregation is still in the hierarchical process of the accretion of its mass. Several facts related with the large scale environment and the faint-end slope of the spectroscopic LF of this group confirm this main result. 

\subsection{The dwarf galaxy population of RXJ075243.6+455653}

Dwarf galaxies in groups and clusters have valuable information about the mass assembly of the systems. There is still an open debate over the formation of the dwarf population in galaxy aggregations. There is some dynamical evidence that at least a fraction of dwarf galaxies residing in galaxy clusters  today were not formed within the cluster \citep[e.g.,][]{conselice2001}. In contrast, they were accreted into the clusters over several Gyr \citep[][]{adami2007, aguerri2017}, or could be produced by the disruption of brighter galaxies due to strong tidal interactions between galaxies or with the cluster potential \citep[e.g.,][]{popesso2006, aguerri2016}. 


The faint-end slope of the galaxy LF indicates the density of the dwarf galaxy population. It depends on many cluster properties such as the dynamical relaxation of the cluster \citep[][]{lopezcruz1997, agulli2017} or the mass of the halo \citep[][]{zandivarez2011}. The LF in groups of galaxies seem to be flatter than those in clusters. \cite{zandivarez2011} found, for a large sample of galaxy groups and clusters from SDSS-DR7, a dependence between the mass of the halo and the faint-end slope of their LFs. In particular, for halos of masses
similar to RXJ075243.6+455653  they obtained $\alpha = -1.16$. The Pearson test indicated that the \cite{zandivarez2011} LF was not a good representation of our data. In contrast, according to this statistical test, the LF of this group is similar to systems showing flatter faint-end slopes in their LFs as those with large magnitude gaps reported by \cite{zarattini2015}. This implies that RXJ075243.6+455653 shows a flatter LF than groups with similar mass, and therefore, the density of dwarfs in RXJ075243.6+455653 is smaller than in groups with similar mass.  

The DGR is a statistic to compare the number of dwarfs per giant galaxy between different clusters or groups. \cite{popesso2006} shows that the DGR measured within $r_{200}$ is independent of the cluster or group mass. We have computed that the DGR of RXJ075243.6+455653 is 0.99 $\pm$ 0.49. This number is one of the smallest values reported by \cite{popesso2006} for a set of groups and clusters. This also points towards RXJ075243.6+455653 having a lower density of dwarf galaxies in comparison with other systems with similar mass. 

Several scenarios could explain the lack of dwarfs observed in RXJ075243.6+455653. It could be related with the accretion status of the system. In particular, the low number of dwarfs in RXJ075243.6+455653 might indicate that this system is still accreting the dwarf galaxy population.  The accretion of the groups in its nearby  environment will increase the number of dwarfs in the system.

RXJ075243.6+455653 has a well defined red sequence down to the dwarf regime. Indeed, all dwarf galaxies of this system  were found to be red  with no detection of blue dwarf galaxies. Galaxy clusters at $z \sim 0.8 - 1.0$ have been found to show a lack of red dwarf galaxies \citep[][]{delucia2007}; and this later formation of the red sequence in the dwarf regime in galaxy clusters indicates that the red dwarf population takes several Gyr to settle down. The fact that RXJ075243.6+455653 already  has red dwarf galaxies could indicate that this group was already formed several Gyr ago, where the red population of dwarf galaxies in the system are likely remnants of early accretion. Numerical simulations show that halos of groups of galaxies with masses similar to RXJ075243.6+455653 have assembled half of their mass at $z \sim 1$ \citep[][]{kundert2017}. Nevertheless, the number of dwarfs of this system is low in comparison with other similar halos, which could indicate that this group is under the hierarchical process of its total mass accretion. The strong dynamical friction suffered by the luminous galaxies
results in their merging with the group brightest galaxy on a short time scale. In contrast, the dynamical friction will be less effective on the dwarf accreted population and will then not merge. This accretion process could decrease the number of luminous galaxies and increase the number of dwarfs. Then, the DGR of the system can increase by the accretion of nearby satellites into the group.

The lack of dwarfs shown by this group can also be explained as a consequence of a destruction of low-mass halos in this system. This galaxy disruption could be due to interactions with the group potential or other galaxy group members \citep[e.g.,][]{popesso2006, aguerri2016}. The low velocity dispersion of the system would favor low-velocity interactions that make stripping of stars from low-mass galaxies more efficient, making them fainter \citep[e.g.,][]{mastropietro2005, aguerri2009, donghia2009, donghia2010}. In contrast, the low mass of the halo of RXJ075243.6+455653 makes it difficult to follow the hypothesis of a massive destruction of dwarf galaxies by halo gravitational shocking \citep[e.g.,][]{mayer2002}. It could also be that dwarf galaxies merge with the dominant galaxy of the system. The low-velocity dispersion of this group would favor this hypothesis. In addition, these minor mergers could  transport some gas to the center of the main galaxy and contribute to the AGN feedback,  which could explain the  bright AGN associated with the central galaxy of RXJ075243.6+455653. Indeed, \cite{hess2012} found that the central galaxy of this group shows extended bipolar jets, as a consequence of this AGN activity.



\subsection{The large-scale environment of fossil systems}

Little is known about the large-scale environment around fossil and non-fossil systems. Only the environment of four fossil systems has been analyzed so far. The results are not conclusive because two of them seem to be embedded in dense large-scale environments and the other two appear isolated \citep[][]{adami2007, adami2012, pierini2011}. Numerical simulations show that FSs are embedded in denser environments at $z \sim 0.4 - 0.7$ than non-FSs. In addition, local FSs are located in slightly more under-dense regions than non-FSs \citep[][]{diazgimenez2011,dariush2010}.  \cite{kundert2017} have shown that the difference between fossil and non-fossil systems identified at $z=0$ is in the mass accretion of their dark matter halos over the past few Gyr. In particular, present-day FSs have assembled 80$\%$ of their mass at higher redshifts than non-FS. This suggests that the large-scale environment might be crucial for the evolution of the magnitude gap. Thus, systems embedded in dense environment could accrete a larger amount of their mass over smaller time-scales than those isolated. This accretion extended along several Gyr would make the fossilness a transitional stage. In addition, extended accretion would also allow for the central galaxies to increase their mass during a larger redshift interval.

The system RXJ075243.6+455653 is embedded in a rich and complex large-scale structure formed by several groups of galaxies. There are 15 small galaxy groups within $\sim$ 7 Mpc around the group and with radial velocities within $V_{c} \pm 3\sigma_{c}$. This indicates that the region around RXJ075243.6+455653 is rich in substructure. In addition, RXJ075243.6+455653 is the dominant structure of this complex system as indicated by the brightest group galaxy located at a peak of X-ray emission with $L_{X} = 2.2 \times 10^{43}\ h_{70}^{-2}$ erg s$^{-1}$ \citep[][]{zarattini2014}. No other surrounding group of galaxies is located at any other X-ray emission peak. This dominance indicates that RXJ075243.6+455653 is surrounded by a massive dark matter halo, and the other nearby galaxy groups will probably fall into this group. This rich environment reveals that RXJ075243.6+455653 is far from the end of the accretion of its mass.

The BGG of RXJ075243.6+455653 has an absolute magnitude of $M_{r} = -22.67$, and is one of the less luminous BGGs in the sample of FSs analyzed by \cite{zarattini2014}. However, this galaxy can still grow by the accretion of the bright galaxies in the nearby environment due to dynamical friction. The galaxies 2 magnitudes fainter than the BGG, located in the galaxy groups, and  with $|V-V_{c}| < 3\sigma_{c}$ , have a total luminosity of $L_{r} = 1.05 \times 10^{12}\ h_{70}^{-2}\ L_{r,\odot}$. All these stars around RXJ075243.6+455653 could potentially produce a central BGG with  $M_{r} =-25.43$. This new galaxy would be comparable in luminosity to the brightest galaxies observed in other fossil clusters and will produce in the future a system with a large magnitude gap \citep[see][]{zarattini2014}. 
Interestingly, this scenario for RXJ075243.6+455653 is in contrast with the prediction from numerical simulations which find that FSs with faint central galaxies are dynamically old objects \citep[see][]{raouf2014}. 

The accretion of galaxies as we propose here for RXJ075243.6+455653 could be not so rare in fossil systems. Recent minor mergers have also been proposed to explain the complex structure of the hot intracluster medium of the fossil group NGC 1132. This system shows a disturbed and asymmetrical morphology of its hot gas. These observations can be explained considering a recent minor merger with a low impact parameter \citep[][]{kim2017}. The case of NGC 1132 shows that fossil systems can be rejuvenated by recent mergers of galaxies or groups. Other X-ray observations indicate that FSs show radial variation of their hot gas content and their gas clumping \citep[][]{pratt2016}. All these observations indicate that fossil systems are not  dynamically evolved systems as was thought.

\section{Conclusions}

 We have analyzed the properties of the fossil group RXJ075243.6+455653. Our main results are as follows:
 
 \begin{itemize}
 \item We have obtained 26 galaxy group members down to $M_{r} = -16.0$ and within $r_{200}$ from the group center. These members provide a mean radial velocity of the group of $V_{c} = 1551 \pm 65$ km s$^{-1}$ and a dispersion $\sigma_{c} = 333 \pm 46$ km s$^{-1}$. 
 \item We have obtained the spectroscopic LF of RXJ075243.6+455653 down to $M_{r} = -16.0$ and within $r_{200}$. The LF was fitted by a single Schechter function giving a flat faint-end slope $\alpha = -1.08 \pm 0.33$. We have used the Pearson test in order to compare the LF of our group with other LFs from the literature. In particular, the Pearson test reports that the LF of RXJ075243.6+455653 is statistically similar to the LF of systems with large magnitude gaps ($\Delta m_{12} > 1.5$) obtained by \cite{zarattini2014}.  And furthermore, the LF of RXJ075243.6+455653 is not statistically similar to the LF of systems with $\Delta m_{12} <0.5$ \citep[][]{zarattini2014} or groups of galaxies with similar mass \citep[][]{zandivarez2011}.
\item The DGR within $r_{200}$ is $0.99\pm0.49$. This value is low compared with the DGR of other groups and galaxy clusters. Both the faint-end of the LF and the DGR of RXJ075243.6+455653 indicate that this group has a lack of faint galaxies compared with groups or clusters with larger $\Delta m_{12}$ or similar halos mass.
 \item Our search of groups or clusters of galaxies in the nearby environment of RXJ075243.6+455653 finds that this group is not isolated. In contrast,  it is embedded in a dense region with up to 15  groups of galaxies located in a thin velocity layer of $\pm 1000$ km s$^{-1}$ and within $\sim 7$ Mpc from the group center. 
 \item RXJ075243.6+455653 is the dominant structure of its complex and rich nearby environment. This is confirmed by the evidence that the central BGG of the group is located at the peak of an extended X-ray emission. No other surrounding galaxy groups are located in any other X-ray peak. This could imply that the surrounding galaxy groups will probably merge with RXJ075243.6+455653 in the next few Gyr.
\end{itemize}

The results obtained in this paper indicate that RXJ075243.6+455653 is still in the process of assembling its mass. The groups of galaxies located in its nearby environment will merge in the next few Gyr. These mergers will produce a growth of the BGG and will increase the number of dwarf galaxies of the system. This indicates that although today RXJ075243.6+455653 is a fossil system, it is not a remnant of an evolved and ancient group. The fossil phase of this system could just be a transition state.  

RXJ075243.6+455653 is a fossil system according to the observational definition of this type of system. Nevertheless, it is not a relic of a structure formed at high redshift, and the halo of this group is still accreting mass from the groups located in its dense nearby environment. RXJ075243.6+455653 is an example showing that the selection of fossil systems by only large magnitude gaps between the first two ranked galaxies does not guarantee the selection of dynamically old galaxy aggregations. New criteria should be obtained in order to find real relics of the structure formation at high redshift. In the near future, we will continue analyzing the properties of the large-scale environment and the faint-end of the LF of other nearby fossil systems. These properties in combination with state-of-the-art cosmological numerical simulations will allow us to find new observables in order to obtain remnants of ancient galaxy aggregations.



\begin{acknowledgements}
      JALA and SZ want to thank the support of this work by the Spanish Ministerio de Economia y Competitividad (MINECO) under the grant AYA2013-43188-P. We thank Dr. R. Barrena for the assistance provided for the data reduction and analysis of the OSIRIS data. This research has made use of the Thirteen Data Release of SDSS, and of the NASA/IPAC Extragalactic Database which is operated by the Jet Propulsion Laboratory, California Institute of Technology, under contract with the National Aeronautics and Space Administration. The WHT and its service program are operated on the island of La Palma by the Isaac Newton Group in the Spanish Observatorio del Roque de los Muchachos of the Instituto de Astrof\'{\i}sica de Canarias. Based on observations made with the Gran Telescopio Canarias (GTC), installed at the Spanish Observatorio del Roque de los Muchachos of the Instituto de Astrof\'{\i}sica de Canarias, in the island of La Palma. This research has made use of the NASA/IPAC Extragalactic Database (NED) which is operated by the Jet Propulsion Laboratory, California Institute of Technology, under contract with the National Aeronautics and Space Administration. 
\end{acknowledgements}

%
%



\end{document}